\newcommand{\CLASSINPUTinnersidemargin}{1 in} 
\documentclass[12pt,journal]{IEEEtran} 
\ifCLASSINFOpdf
\else
\fi

\usepackage{color}
\usepackage{amssymb}
\usepackage{amsthm}
\usepackage[cmex10]{amsmath}
\DeclareMathOperator{\E}{\mathbb{E}}

\DeclareMathOperator{\C}{\mathbb{C}}

\newcommand {\Define} {\stackrel {\Delta} {=}  }

\newcommand{\mya}{\mathrel{\overset{\makebox[0pt]{{\tiny(a)}}}{=}}}

\newcommand{\myb}{\mathrel{\overset{\makebox[0pt]{{\tiny(b)}}}{=}}}

\newcommand{\myc}{\mathrel{\overset{\makebox[0pt]{{\tiny(c)}}}{=}}}

\usepackage{bm}
\usepackage{mathtools}
\usepackage{fixltx2e}
\usepackage{epsfig,makeidx,color}
\usepackage{graphicx,dblfloatfix}
\usepackage{amsbsy}
\usepackage{amssymb}
\usepackage{euscript}
\usepackage{lipsum}
\usepackage{cite}
\usepackage{chngcntr}
\usepackage{lineno}
\usepackage[T1]{fontenc}
\usepackage{microtype}
\usepackage{wasysym}
\usepackage{footnote}
\usepackage[left=01 in,top=0.5 in,right= 1 in,bottom=0.5 in]{geometry}        
\newtheorem{theorem}{Theorem}
\newtheorem{lemma}{Lemma}
\newtheorem{proposition}{Proposition}

\newtheorem{remark}{\it Remark}

\newtheorem{result}{\it Result}

\usepackage[english]{babel}
    \usepackage[font=small,labelsep=space]{caption}
\usepackage{subcaption}    
\renewcommand{\baselinestretch}{1.54}
\makeatletter
\def\citenoauxwrite#1{\begingroup
\@fileswfalse
\cite{#1}\relax
\endgroup}
\makeatother

\begin{document}

\title{Impact of CFO Estimation on the Performance of ZF Receiver in Massive MU-MIMO Systems}
%
%
%

\author{Sudarshan~Mukherjee, 
        ~Saif Khan~Mohammed and Indra Bhushan
\thanks{The authors are with the Department of Electrical Engineering, Indian Institute of Technology (I.I.T.) Delhi, India. Saif Khan Mohammed is also associated with Bharti School of Telecommunication Technology and Management (BSTTM), I.I.T. Delhi. Email: saifkmohammed@gmail.com. This work is supported by EMR funding from the Science and Engineering
Research Board (SERB), Department of Science and Technology (DST),
Government of India.}
}

\onecolumn
\maketitle

\vspace{-2.5 cm}

\begin{abstract}
In this paper, we study the impact of carrier frequency offset (CFO) estimation/compensation on the information rate performance of the zero-forcing (ZF) receiver in the uplink of a multi-user massive multiple-input multiple-output (MIMO) system. Analysis of the derived closed-form expression of the per-user information rate reveals that with increasing number of BS antennas $M$, an $\mathcal{O}(\sqrt{M})$ array gain is achievable, which is same as that achieved in the ideal zero CFO scenario. Also it is observed that compared to the ideal zero CFO case, the performance degradation in the presence of residual CFO (after CFO compensation) is the same for both ZF and MRC.
\end{abstract}

\vspace{-0.8 cm}
\begin{IEEEkeywords}

\vspace{-0.4 cm}
Massive MIMO, carrier frequency offset (CFO), multi-user, array gain, zero-forcing (ZF).
\end{IEEEkeywords}

\vspace{-0.5 cm}

%

\vspace{-0.4 cm}

\section{Introduction}
%
%
%
%
 In the past few years, massive multiple-input multiple-output (MIMO) systems have emerged as one of the key technologies in the evolution of the next generation $5$G wireless systems due to their ability to support high data rate and improved energy efficiency  \cite{Andrews, Boccardi}. In a massive multi-user (MU) MIMO system, the base station (BS) is provided with hundreds of antennas to simultaneously serve only a few tens of single-antenna user terminals (UTs) in the same time-frequency resource \cite{Marzetta1}. Increasing the number of BS antennas open up more available degrees of freedom, which helps accommodate more number of users, thus improving the achievable spectral efficiency \cite{Larsson, Marzetta2}. At the same time, the required radiated power to achieve a fixed desired information rate can be reduced with increasing number of BS antennas, $M$ (array gain). It has been shown that even with imperfect channel state information (CSI), the achievable array gain for any sub-optimal linear receiver (e.g. zero-forcing (ZF), maximum ratio combining (MRC) etc.) is $\mathcal{O}(\sqrt{M})$ \cite{Ngo1}. 
 
 \par Above results assume perfect frequency synchronization at the BS receiver, without which the performance of the system would deteriorate rapidly. In practice acquiring perfect knowledge of the carrier frequency offsets (CFOs) between the received user signals at the BS and the frequency of the BS oscillator is however a challenging task. There exists various techniques for CFO estimation and compensation for conventional small MIMO systems in the literature \cite{Ma,Simon, Ghogho, Poor}. However these algorithms incur tremendous increase in computational complexity with increasing number of BS antennas, $M$ and increasing number of UTs, $K$ (i.e. massive MIMO scenario). Recently in \cite{Larsson2} an approximation to the joint ML (Maximum Likelihood) CFO estimation has been proposed for massive MIMO system. However this technique requires a multi-dimensional grid search and therefore has high complexity with large number of UTs.
 
 \par {In \cite{gcom2015} the authors propose a simple low complexity algorithm for CFO estimation and a corresponding communication strategy for massive MU-MIMO uplink. It has been shown that with sufficiently large $M$, the algorithm has only $\mathcal{O}(M)$ complexity (independent of the number of UTs). However the impact of the residual CFO (due to CFO compensation) on the performance of massive MIMO is yet to be studied. The most common linear suboptimal receivers used in massive MIMO uplink are MRC (maximum ratio combining) and ZF (zero-forcing) receivers. With the MRC receiver, system performance is limited by the multi-user interference (MUI) in the high SNR regime}. For the ideal zero CFO scenario, the ZF receiver is known to remove this limitation by eliminating the MUI \cite{Ngo1}. In this work we therefore study the impact of the residual CFO error (due to the CFO estimation strategy proposed in \cite{gcom2015}) on the achievable information rate of the ZF receiver and compare it to that of the MRC receiver. To the best of our knowledge, this paper is the first to report such a study.
 
 \par The contributions of our paper are as follows: (i) we have derived a closed-form expression for an achievable information rate for the ZF receiver with MMSE (minimum mean square error) channel estimation and CFO compensation. A closed-form expression for the same is also derived for MRC; (ii) analysis of the ZF information rate expression reveals that an $\mathcal{O}(\sqrt{M})$ array gain is achievable. This is very interesting since even for the ideal zero CFO scenario, the best possible array gain is known to be $\mathcal{O}(\sqrt{M})$ only \cite{Ngo1}; (iii) for the same desired per-user information rate, the SNR gap (i.e. the extra SNR required by MRC when compared to ZF) does not degrade with CFO estimation/compensation, when compared to the ideal zero CFO case. This suggests that compared to the ideal zero CFO case, the performance degradation in the presence of residual CFO (due to compensation) is the same for both ZF and MRC. [\textbf{{Notations:}} $\C$ denotes the set of complex numbers. $\E$ denotes the expectation operator. $(.)^{H}$ denotes the complex conjugate transpose operation, while $(.)^{\ast}$ denotes the complex conjugate operator. Also, $\bm I_{N}$ denotes the $N\times N$ identity matrix and $\bm A_{mk}$ (or $(\bm A)_{mk}$) denotes the $(m,k)$-th element of matrix $\bm A$.]

\vspace{-0.5 cm}

\section{System Model}

\begin{figure}[t]
\vspace{-0.7 cm}
\centering
\includegraphics[width= 5.6 in, height= 1.6 in]{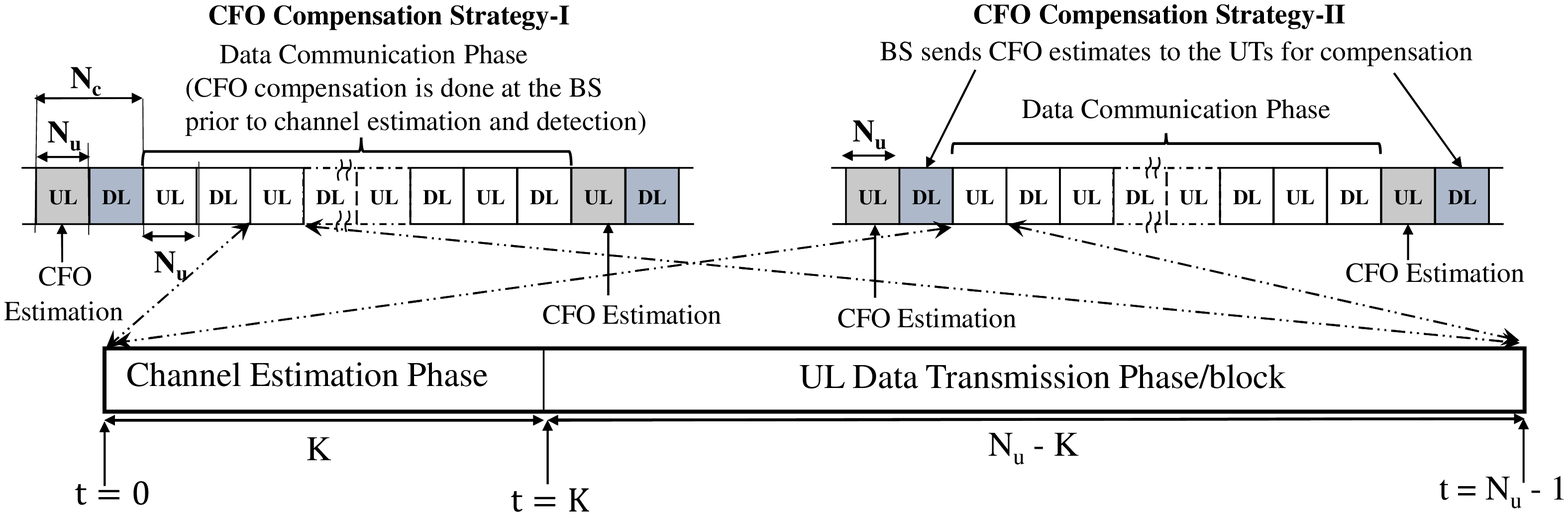}
\caption {The communication strategy: CFO Estimation and Compensation Strategies and Data Communication.} 
\label{fig:uplink}
\vspace{-0.8 cm}
\end{figure}

We consider a frequency-flat massive MU-MIMO uplink (UL) channel, where the massive MIMO BS is equipped with $M$ BS antennas and is coherently communicating with $K$ single antenna UTs simultaneously in the same time-frequency resource. Therefore for a massive MU-MIMO BS, acquisition and compensation of CFOs from different UTs is important. Since in massive MIMO, the BS is expected to operate in time division duplexed (TDD) mode, the coherence interval ({of $N_c$ channel uses}) consists of a UL slot ({$N_u$ channel uses}), followed by a downlink (DL) slot ({$N_c - N_u$ channel uses}). {As for the communication strategy (see in Fig.~\ref{fig:uplink}), we perform CFO estimation in a special UL slot prior to the data communication. For CFO estimation, we adopt the CFO estimation strategy presented in \cite{gcom2015}. CFO compensation can be performed in two different ways -- (i) at the BS (prior to channel estimation and multi-user detection); or (ii) at the respective UTs prior to data transmission (this however requires transmission of CFO estimates from the BS to the UTs over a control channel in the DL slot, following the special UL slot for CFO estimation).} Data communication starts from the first UL slot, following the special CFO estimation UL slot. In these UL slots, prior to UL data transmission, the UTs transmit pilots for channel estimation.{\footnote[1]{{The same CFO estimates can be used for CFO compensation prior to precoding in the DL slot of each coherence interval.}}} The special UL/DL slot for CFO estimation might be repeated every few coherence intervals, depending on how fast the CFOs change.
\vspace{-0.6 cm}

\subsection{CFO Estimation Strategy in \cite{gcom2015}}

For the CFO estimation phase, special pilots are transmitted by the UTs in the uplink. {A pilot sequence of length $N \leq N_u$ is divided into $B = N/K$ pilot-blocks, where each pilot-block is $K$ channel uses long.\footnote[2]{{Although the CFO estimation method assumes $N/K$ to be integer, we can accommodate non-integer values of $N/K$, by defining the number of blocks $B \Define \lceil N/K \rceil$. Hence for non-integer $N/K$, the $B^{\text{th}}$ block is less than $K$ channel uses long. Therefore the effective $N/K$ is $\lceil N/K \rceil$ for UTs allowed to transmit in the $B^{\text{th}}$ block, and it is $(\lceil N/K \rceil - 1)$ for all other UTs.}}} Each UT transmits only a single impulse of amplitude $\sqrt{K p_{\text{u}}}$ in each pilot-block. Therefore in the $b^{\text{th}}$ pilot-block, the $k^{\text{th}}$ UT transmits impulse at $t = \tau(b,k) = (b-1)K + k-1$, where $k = 1, 2, \ldots, K$ and $b = 1, 2, \ldots, B$. The pilot signal received at the $m^{\text{th}}$ BS antenna at time $\tau(b,k)$ is therefore given by $r_m[\tau(b,k)] = \sqrt{K p_{\text{u}}} \,\, g_{mk} \, e^{j\omega_k \tau(b,k)} + w_m[\tau(b,k)]$, where $\omega_k \Define 2\pi \Delta f_k T_s$ is the CFO for the $k^{\text{th}}$ user ($T_s = 1/B_c$, where $B_c$ is the communication bandwidth and $\Delta f_k$ is the frequency offset of the $k^{\text{th}}$ UT). Also $g_{mk} \Define h_{mk}\sqrt{\beta_k}, m = 1, 2, \ldots, M$ and $k = 1, 2, \ldots, K$, is independent complex baseband frequency-flat channel gain coefficient between the $m^{\text{th}}$ BS antenna and the $k^{\text{th}}$ UT and $h_{mk} \sim \mathcal{C}\mathcal{N}(0,1)$.{\footnote[3]{{Independent and identically distributed Rayleigh fading is a commonly used model for the distribution of channel gains in a massive MIMO system \cite{Marzetta1, Larsson, Marzetta2}.}}} $\sqrt{\beta_k} > 0$ models the geometric attenuation factor for the $k^{\text{th}}$ user and $w_m[\tau(b,k)] \sim \mathcal{C}\mathcal{N}(0, \sigma^2)$ is the complex circular symmetric AGWN noise with variance $\sigma^2$. The estimate of the CFO of the $k^{\text{th}}$ UT, $\widehat{\omega}_k$, is obtained as the principal argument of the block-wise correlation term of the pilot sequence received from the $k^{\text{th}}$ user, i.e., $\widehat{\omega}_k = \frac{1}{K}\arg{(\rho_k)}$,\footnote[4]{Here $\arg{(c)}$ denotes the \lq{principal argument}' of the complex number $c$.} where\footnote[5]{For expression of $\nu_k$ see \cite{gcom2015}.} \small{$\rho_k \Define \dfrac{\sum\limits_{b=1}^{B-1} \sum\limits_{m=1}^{M} r_m^{\ast}[\tau(b,k)]r_m[\tau(b+1,k)]}{M K (B-1) p_{\text{u}} \beta_k} =  G_k e^{j\omega_k K} + \nu_k$}\normalsize, and $G_k \Define \dfrac{1}{M}\sum\limits_{m = 1}^{M} |h_{mk}|^2$.

\begin{remark}
\normalfont
Note that the above CFO estimate is well-defined iff $|\omega_k K| < \pi$. For most practical massive MIMO systems, this condition will hold true \cite{gcom2015}. Also, from the strong law of large numbers it can be shown that for i.i.d. $h_{mk}$, $G_k \to 1$ as $M \to \infty$. \hfill \qed
\end{remark}

\begin{result}
\normalfont
(\emph{Approximation of the CFO Estimate in \cite{gcom2015}}): If $|\omega_k K| \ll \pi$ and $\gamma \Define \frac{p_{\text{u}}}{\sigma^2} \gg \gamma^{0}$, then the above CFO estimate can be approximated by {$\widehat{\omega}_k  = \frac{1}{K}\arg{(\rho_k)} \approx  \omega_k + \frac{\nu_k^Q}{G_k K}$}, where \small{$\gamma^{0} \Define \dfrac{\frac{B-1}{2B-3}}{KG_k\bigg[\sqrt{1 + 2M\frac{(B-1)^3}{(2B-3)^2}} - 1\bigg]}$}\normalsize and $\nu_k^Q \Define \Im (\nu_k)$. Note that $(\widehat{\omega}_k - \omega_k) \sim \mathcal{N}(0, \sigma_{\omega_k}^2)$, where $\sigma_{\omega_k}^2$ is the mean square error (MSE) given by 

\vspace{-1.1 cm}

\small{\begin{IEEEeqnarray}{rCl}
\label{eq:cfomse}
\sigma_{\omega_k}^2 \Define \E[(\widehat{\omega}_k - \omega_k)^2] & \approx & \dfrac{\frac{1}{\gamma \beta_k}\left(\frac{G_k}{B-1} + \frac{1}{2K\gamma \beta_k}\right)}{M(N-K)K^2G_k^2}.
\end{IEEEeqnarray}}\normalsize
\end{result}

\vspace{-0.5 cm}

\begin{remark}
\normalfont
Clearly with $M \to \infty$, we have $\gamma^{0} \propto \frac{1}{\sqrt{M}}$. Therefore we choose some constant $c_0 > 0$ such that $\gamma = \frac{c_0}{\sqrt{M}} \gg \gamma^{0}$ as $M \to \infty$, thereby satisfying the required condition $\gamma \gg \gamma^{0}$ in \textit{Result} $1$. From \eqref{eq:cfomse} we note that with $\gamma = \frac{c_0}{\sqrt{M}}$ and $M \to \infty$ (fixed $N$, $K$), we have \small{$\E[(\widehat{\omega}_k - \omega_k)^2] \approx \dfrac{1/c_0^2}{2K^3(N-K)\beta_k^2}$}\normalsize, since $\lim\limits_{M \to \infty} G_k = 1$. This shows that the MSE for CFO estimation approaches a constant value as $M \to \infty$ with $\gamma \propto 1/\sqrt{M}$ for fixed $K$ and fixed $N$. {Note that with $M \to \infty$, the desired MSE $\propto 1/c_0^2$, i.e., a smaller desired MSE can be attained using a higher value of $c_0$. Therefore for a target/desired MSE, a sufficiently large $M$ must be chosen so that the required power ($\propto 1/\sqrt{M}$) for CFO estimation is within the desired limits.} \hfill \qed
\end{remark}

\vspace{-0.7 cm}

\subsection{Uplink Data Communication}

{After CFO estimation, CFO compensation can be performed in one of the following two ways: (a) the BS can feed the individual CFO estimates back to the corresponding UTs over a control channel in the DL slot, following the special UL slot (see Fig.~\ref{fig:uplink}). In this way, the $k^{\text{th}}$ UT would correct its CFO by rotating the transmit signal at the $t^{\text{th}}$ channel use by $e^{-j \widehat{\omega}_k t}$. However in this method there is possibility of corruption of the estimates due to error in the control channel; (b) another way of correcting frequency offsets is to perform CFO compensation at the BS, prior to channel estimation and multi-user detection. In this paper, we would use this second technique for CFO compensation and study the impact of the residual CFOs (i.e. $\widehat{\omega}_k - \omega_k$) on the performance of massive MIMO uplink.\footnote[6]{{Note that the information theoretic performance is identical for both the CFO compensation techniques.}}} The uplink data communication starts at $t = 0$ (see Fig.~\ref{fig:uplink}). We assume that in the first $K$ consecutive channel uses, the UTs transmit pilots for channel estimation sequentially in time{\footnote[7]{{Though impulse type pilots are not amenable to practical implementation (due to high peak-to-average-power ratio (PAR)), we use them because our main objective is to study the first order effects of system parameters $M$, $K$, $N$, $p_{\text{u}}$ etc. on the information rate performance of ZF and MRC detectors in the presence of residual CFOs.}}}, i.e., the $k^{\text{th}}$ UT transmits an impulse of amplitude $\sqrt{K \, p_{\text{u}}}$ only in the $(k-1)^{\text{th}}$ channel use. {The received pilot at the $m^{\text{th}}$ BS antenna at $t = k-1$ is therefore given by $r_m[k-1] = \sqrt{K \, p_{\text{u}}} \, g_{mk}\, e^{j \omega_k (k-1)} + w_m[k-1]$, where $k = 1, 2, \ldots, K$ and $m = 1, 2, \ldots, M$. $w_m[k-1] \sim \mathcal{C}\mathcal{N}(0,\sigma^2)$ is the circular symmetric AWGN.}

\vspace{-0.61 cm}

\subsection{MMSE Channel Estimation}

{To estimate the channel gains, firstly, CFO compensation is performed on the received pilots at the BS. Since the pilots from different UTs are separated in time, pilot from the $k^{\text{th}}$ UT after compensation is given by $y_m[k-1] = r_m[k-1]e^{-j\widehat{\omega}_k (k-1)} = \sqrt{K \, p_{\text{u}}}\,\, \widetilde{g}_{mk} + n_{mk}[k-1]$, where\footnote[8]{{Both $g_{mk}$ and $w_m[k-1]$ have uniform phase distribution (i.e. circular symmetric) and are independent of each other. Clearly, rotating these random variables by fixed angles (for a given realization of CFOs and its estimates) would not change the distribution of their phases and they will remain independent. Therefore the distribution of $\widetilde{g}_{mk}$ and $n_{mk}[k-1]$ would be same as that of $g_{mk}$ and $w_m[k-1]$ respectively.}} $\widetilde{g}_{mk} \Define g_{mk}\, e^{-j \Delta \omega_k (k-1)} \sim \mathcal{C}\mathcal{N}(0,1)$, $n_{mk}[k-1] \Define w_m[k-1] e^{-j\widehat{\omega}_k (k-1)} \sim \mathcal{C}\mathcal{N}(0, \sigma^2)$ and $\Delta \omega_k \Define \widehat{\omega}_k - \omega_k$ is the residual CFO error. Next we compute the minimum mean square estimate (MMSE) of the effective channel gain coefficient $\widetilde{g}_{mk}$ as given below:}

\vspace{-0.8 cm}

\small{\begin{IEEEeqnarray}{rCl}
\label{eq:chanestmk}
\widehat{g}_{mk} = \frac{\sqrt{K p_{\text{u}}} \beta_k}{K p_{\text{u}}\beta_k + \sigma^2}y_m[k-1] = \frac{\sqrt{K p_{\text{u}}} \beta_k}{K p_{\text{u}}\beta_k + \sigma^2} \Big(\sqrt{K p_{\text{u}}} \, g_{mk} \, e^{-j\Delta \omega_k (k-1)}  + n_{mk}[k-1]\Big).
\IEEEeqnarraynumspace
\end{IEEEeqnarray}}\normalsize

\vspace{-0.3 cm}

\noindent  where $m = 1, 2, \ldots, M$ and $k = 1, 2, \ldots, K$. Using \eqref{eq:chanestmk}, the estimate of the effective channel gain matrix is given by

{\vspace{-1.4 cm}

\begin{IEEEeqnarray}{rCl}
\label{eq:chanest}
\widehat{\bm G} = (\sqrt{K p_{\text{u}}} \bm G \bm \Phi_0 + \bm N) \widetilde{\bm D},
\IEEEeqnarraynumspace
\end{IEEEeqnarray}}

\vspace{-1.1 cm}

\noindent where $\widehat{\bm G} \Define [\widehat{g}_{mk}]_{M \times K}$, $\bm G \Define [g_{mk}]_{M \times K}$, and $\bm \Phi_0 \Define diag(1, e^{-j \Delta\omega_2}, \cdots, e^{-j \Delta \omega_K (K-1)})$. Here $\bm N = [n_{mk}[k-1]]_{M \times K}$ and $\widetilde{\bm D} \Define (\sqrt{K p_{\text{u}}} \bm I_K + \frac{\sigma^2}{\sqrt{K p_{\text{u}}}}\bm D^{-1})^{-1}$, where $\bm D \Define diag(\beta_1, \beta_2, \cdots, \beta_K)$.

\vspace{-0.5 cm}

\section{Uplink Receiver Processing}

In this section we formulate a generalized approach towards multi-user receiver processing at the massive MIMO BS. From Fig.~\ref{fig:uplink} it is clear that uplink data transmission begins at the $t = K^{\text{th}}$ channel use and continues till the $(N_u - 1)^{\text{th}}$ channel use, where $N_u$ is the duration of the UL slot. Let $\sqrt{p_{\text{u}}} \, x_k[t]$ be the information symbol transmitted by the $k^{\text{th}}$ UT at the $t^{\text{th}}$ channel use. {The signal received at the $m^{\text{th}}$ BS antenna in the $t^{\text{th}}$ channel use is given by $r_m[t] = \sqrt{p_{\text{u}}} \sum\limits_{q=1}^{K} g_{mq} \, e^{j \omega_q t}\, x_q[t] + w_m[t]$, where $m = 1, 2, \ldots, M$. To detect information symbols of the $k^{\text{th}}$ UT, first, CFO compensation is performed, followed by detection using the detector for the $k^{\text{th}}$ UT. In this paper, we only consider linear detectors.} We also assume $x_k[t] \sim \mathcal{C}\mathcal{N}(0,1)$, for $t = K, \cdots, N_u - 1$ and are i.i.d. Let $\bm r[t] \Define (r_1[t], r_2[t], \cdots, r_M[t])^T$. The detected signal from the $k^{\text{th}}$ UT at the BS after CFO compensation is given by

\vspace{-0.8 cm}

{\small{\begin{IEEEeqnarray}{lCl}
\label{eq:xhat}
\nonumber \widehat{x}_k [t] & = & \underbrace{\bm a_k^H \bm r[t]}_{\text{Detection}} \, \underbrace{e^{-j\widehat{\omega}_k t}}_{\stackrel{\text{CFO}}{\text{compensation}}} = \sum\limits_{m = 1}^{M} a_{mk}^{\ast} r_m[t]\, e^{-j\widehat{\omega}_k t} = \sqrt{p_{\text{u}}} \sum\limits_{q = 1}^{K} \Bigg( \underbrace{\sum\limits_{m = 1}^{M} a_{mk}^{\ast}\, g_{mq}}_{ \Define \bm a_k^H \bm g_q} \, e^{j (\omega_q - \widehat{\omega}_k) t}\, x_q[t] \Bigg) + \underbrace{\sum\limits_{m = 1}^{M} a_{mk}^{\ast}\, n_{mk}[t]}_{ \Define \bm a_k^H \bm n_k[t]}\\
\nonumber & = & \sqrt{p_{\text{u}}} \bm a_k^H \bm g_k e^{-j \Delta \omega_k t}\, x_k[t] + \sqrt{p_{\text{u}}} \sum\limits_{q = 1, q \neq k}^{K} \bm a_k^H \bm g_q e^{-j \Delta \omega_q t}\, x_q[t]\, e^{j(\widehat{\omega}_q - \widehat{\omega}_k)t} + \bm a_k^H \bm n_k[t]\\
& = & \sqrt{p_{\text{u}}} \bm a_k^H  \widetilde{\bm g}_k e^{-j \Delta \omega_k (t - (k-1))}\, x_k[t] + \sqrt{p_{\text{u}}} \sum\limits_{q = 1, q \neq k}^{K} \bm a_k^H \widetilde{\bm g}_q e^{-j \Delta \omega_q (t - (q-1))}\, x_q[t]\, e^{j(\widehat{\omega}_q - \widehat{\omega}_k)t} + \bm a_k^H \bm n_k[t],
\IEEEeqnarraynumspace
\end{IEEEeqnarray}}\normalsize}

\vspace{-0.9 cm}

\noindent where $\bm a_k \Define (a_{1k}, a_{2k}, \cdots, a_{Mk})^T \in \C^{M \times 1}$ is the linear detector for the $k^{\text{th}}$ user, $\Delta \omega_k = \widehat{\omega}_k - \omega_k$ and {$n_{mk}[t] \Define w_m[t] e^{-j \widehat{\omega}_k t}$. Also, $\bm n_k[t] \Define (n_{1k}[t], n_{2k}[t], \cdots, n_{Mk}[t])^T$}, $\bm g_q \Define (g_{1q}, g_{2q}, \cdots, g_{Mq})^T$ (the $q^{\text{th}}$ column of $\bm G$) {and $\widetilde{\bm g}_q \Define (\widetilde{g}_{1q}, \widetilde{g}_{2q}, \cdots, \widetilde{g}_{Mq})^T = \bm g_q \, e^{-j \Delta \omega_q (q-1)}$.} 

\vspace{-0.6 cm}

\subsection{Coding Strategy}

{We define the effective channel estimation error as $\epsilon_{mk} \Define \widehat{g}_{mk} - \widetilde{g}_{mk}$ (see Section II-C). Let $\bm \epsilon_k \Define (\epsilon_{1k}, \epsilon_{2k}, \cdots, \epsilon_{Mk})^T = \widehat{\bm g}_k - \widetilde{\bm g}_k$, where $\widehat{\bm g}_k = (\widehat{g}_{1k}, \widehat{g}_{2k}, \cdots, \widehat{g}_{Mk})^T$ ($k^{\text{th}}$ column of $\widehat{\bm G}$). The mean vector and the covariance matrix of $\bm \epsilon_k$ are respectively given by $\E[\bm \epsilon_k] = \bm 0$ and} 

\vspace{-0.8 cm}
{\small{\begin{IEEEeqnarray}{rCl}
\label{eq:errvar}
\E[\bm \epsilon_k \bm \epsilon_k^H] & = &  \E \Bigg[\Bigg( \underbrace{\left(\dfrac{{K p_{\text{u}}} \beta_k}{K p_{\text{u}} \beta_k + \sigma^2} - 1 \right)\widetilde{\bm g}_k + \dfrac{\sqrt{K p_{\text{u}}} \beta_k}{K p_{\text{u}} \beta_k + \sigma^2}\bm n_k[k-1]}_{\widehat{\bm g}_k - \widetilde{\bm g}_k} \Bigg) \Big(\widehat{\bm g}_k - \widetilde{\bm g}_k\Big)^H \Bigg] \mya \dfrac{\beta_k \, \sigma^2}{K p_{\text{u}} \beta_k + \sigma^2}  \bm I_M,
\IEEEeqnarraynumspace
\end{IEEEeqnarray}}\normalsize}

\vspace{-0.5 cm}

{\noindent where $(a)$ follows from \eqref{eq:chanestmk} and $\bm n_k[k-1] = (n_{1k}[k-1], \cdots, n_{Mk}[k-1])^T$. Using $\widetilde{\bm g}_k = \widehat{\bm g}_k - \bm \epsilon_k$ in \eqref{eq:xhat}, we get}

\vspace{-1 cm}

{\small{\begin{IEEEeqnarray}{rCl}
\label{eq:modifyxhat}
\nonumber \widehat{x}_k[t] & = & \underbrace{\sqrt{p_{\text{u}}} \bm a_k^H \widehat{\bm g}_k \, e^{-j \Delta \omega_k (t - (k-1))}}_{\Define \,\, S_k[t]} \, x_k[t] + \underbrace{\bm a_k^H \bm n_k[t]}_{\Define \,\, \text{EN}_k[t]}\\
\nonumber & &  + \underbrace{\sqrt{p_{\text{u}}} \Bigg( \sum\limits_{\substack{q = 1,\\q \neq k}}^{K}  \Big(\bm a_k^H \Big(\widehat{\bm g}_q - \bm \epsilon_q \Big) e^{-j \Delta \omega_q (t - (q-1))}\, x_q[t]\Big)\, e^{j(\widehat{\omega}_q - \widehat{\omega}_k)t} - \bm a_k^H \bm \epsilon_k e^{-j \Delta \omega_k (t - (k-1))}\, x_k[t]\Bigg)}_{\Define \,\, \text{MUI}_k[t]}\\
& = &  \underbrace{\E \Big[S_k[t] \Big] x_k[t]}_{\Define \,\, \text{ES}_k[t]} + \underbrace{\left(S_k[t] - \E\Big[S_k[t]\Big]\right) x_k[t]}_{\Define \,\, \text{SIF}_k[t]}\, + \text{MUI}_k[t] + \text{EN}_k [t] ,
\IEEEeqnarraynumspace
\end{IEEEeqnarray}}\normalsize}

\vspace{-0.5 cm}

\noindent where $\text{SIF}_k[t]$ is the time-varying self-interfering component of the desired signal and $\text{SIF}_k [t] + \text{MUI}_k[t] + \text{EN}_k [t] \Define W_k[t]$ is the overall effective noise term. Further $\E \Big[S_k[t] \Big]$ is the average value of $S_k[t]$, across several uplink data transmission blocks, i.e., several channel realizations, and is a function of $t$. The same is also true for the variance of $W_k[t]$. Furthermore for a given $t$, across multiple uplink data transmission blocks, the realizations of $W_k[t]$ are i.i.d. Hence for each channel use, $t = K, K+1, \ldots, N_u - 1$, we have a additive noise SISO (single-input single-output) channel in \eqref{eq:modifyxhat}. Thus for each user there are $N_u -K$ different SISO channels with distinct channel statistics. Therefore we consider $N_u - K$ channel codes for each user, one for each SISO channel. The data received for each user in the $t^{\text{th}}$ channel use across multiple coherence intervals is jointly decoded at the receiver \cite{Phasenoise}. This coding strategy albeit not practical is useful in computing an achievable information rate.\footnote[9]{In practice, coding could be performed across a group of consecutive channel uses within each transmission block, since the statistics of $W_k[t]$ and $S_k[t]$ would not change significantly within a small group of consecutive channel uses.}

\vspace{-0.7 cm}

\section{Achievable Information Rate}

\vspace{-0.3 cm}

In essence, from the above coding strategy, we have $N_u - K$ parallel channel decoders for each user. For the $t^{\text{th}}$ SISO channel of the $k^{\text{th}}$ user, we note that the correlation between the desired signal term $\text{ES}_k[t]$ and the overall effective noise $W_k[t]$ is zero, i.e., from \eqref{eq:modifyxhat} we have

\vspace{-0.9 cm}

{\small{\begin{IEEEeqnarray}{lCl}
\label{eq:croscor}
\nonumber \E \Big[\text{ES}_k^{\ast}[t] W_k[t]\Big] \mya \E \Big[S_k^{\ast}[t]\Big] \E \Bigg[\underbrace{|x_k[t]|^2 \Big\{ S_k[t] - \E \Big[S_k[t]\Big]  \Big\}}_{= \, 0, \, \text{since $x_k[t]$ and $S_k[t]$ are independent}} + x_k^{\ast}[t]\text{MUI}_k[t] + \underbrace{x_k^{\ast}[t] \text{EN}_k[t]}_{\substack{= \,\, 0,\, \text{since}\, \bm n_k[t]\, \text{is zero mean and}\\ \text{independent of} \,x_k[t]}}\Bigg]\\
 \myb \E \Big[S_k^{\ast}[t]\Big] \E \Bigg[\sqrt{p_{\text{u}}} \Bigg(\underbrace{\stackrel{x_k^{\ast}[t]  \sum\limits_{{q = 1, q \neq k}}^{K} \Big(\bm a_k^H (\widehat{\bm g}_q - \bm \epsilon_q) e^{-j \Delta \omega_q (t - (q-1))}\, x_q[t]\Big)\,}{\times e^{j(\widehat{\omega}_q - \widehat{\omega}_k)t}}}_{= \, 0, \, \text{since} \, x_i[t] \, \text{are all i.i.d.}} - \underbrace{\bm a_k^H \bm \epsilon_k e^{-j \Delta \omega_k (t - (k-1))}\, |x_k[t]|^2}_{\substack{= \, 0, \, \text{since} \,\, \widehat{\bm g}_k \text{and} \,\, \bm \epsilon_k \text{are orthogonal due to}\\ \text{ MMSE estimation and } \bm a_k \, \text{is function of} \, \widehat{\bm g}_k}}\Bigg)\Bigg]  =  0.
\IEEEeqnarraynumspace
\end{IEEEeqnarray}}\normalsize}

\vspace{-0.7 cm}

 \noindent where $(a)$ and $(b)$ follow from the definitions of $\text{SIF}_k[t]$ and $\text{MUI}_k[t]$ in \eqref{eq:modifyxhat}. With Gaussian information symbols $x_k[t]$, a lower bound on the information rate of the effective channel in \eqref{eq:modifyxhat} is obtained by considering the \emph{worst case uncorrelated additive noise} (in terms of mutual information), having the same variance as $W_k[t]$. With Gaussian information symbols, this worst case uncorrelated noise is also Gaussian \cite{Hasibi2}. The variance of $W_k[t]$ is given by $\E[|W_k[t]|^2] = \E[|\text{SIF}_k[t] + \text{MUI}_k[t] + \text{EN}_k [t]|^2]$. Since all $x_k[t]$ and $\bm n_k[t]$ are independent and zero mean, it can be shown that $\E[\text{SIF}_k^{\ast}[t] \text{EN}_k [t]] = \E[\text{MUI}_k^{\ast}[t] \text{EN}_k [t]] = 0$. Also due to MMSE channel estimate it can be shown that $\E[\text{SIF}_k^{\ast}[t] \text{MUI}_k[t]] = 0$. Therefore $\E[|W_k[t]|^2] = \E[|\text{SIF}_k[t]|^2] + \E[|\text{MUI}_k[t]|^2] + \E[|\text{EN}_k [t]|^2]$. Also, $\E\Big[\text{ES}_k[t]\Big] = \E\Big[\text{SIF}_k[t]\Big] = \E\Big[\text{MUI}_k[t]\Big] = \E\Big[\text{EN}_k[t]\Big] = 0$. An achievable rate is therefore given by the following lower bound on $I(\widehat{x}_k[t]; x_k[t])$
 
 \vspace{-1.1 cm}

{\begin{IEEEeqnarray}{rCl}
\label{eq:infolb}
I(\widehat{x}_k[t]; x_k[t]) \geq \log_2(1 + \text{SINR}_k [t]), \,\, \text{where $\text{SINR}_k[t] \Define {\E \Big[|\text{ES}_k [t]|^2\Big]}\Big /{\E \Big[|W_k[t]|^2 \Big]}$},
\IEEEeqnarraynumspace
\end{IEEEeqnarray}}\normalsize

\vspace{-0.3 cm}

\noindent and the overall information rate for the $k^{\text{th}}$ user is thus given by{\footnote[10]{{In a wireless channel of bandwidth 200 KHz and a coherence interval of duration 1 millisecond, even with $K = 10$ UTs, the channel estimation overhead is only 5$\%$. Further, CFO estimation is performed at a 5 to 10 times slower rate than channel estimation and therefore its overhead is expected to be less than 1$\%$ \cite{Skold}. We have therefore neglected the CFO estimation overhead in \eqref{eq:sumrate}, since it is a mere scaling factor, which does not impact the main conclusions of our work.}}}

\vspace{-1.05 cm}

{\begin{IEEEeqnarray}{rCl}
\label{eq:sumrate}
I_k = \frac{1}{N_u} \sum\limits_{t = K}^{N_u - 1} \log_2 (1 + \text{SINR}_k [t]).
\end{IEEEeqnarray}}\normalsize

\vspace{-0.8 cm}

\subsection{Mutual Information Analysis for the ZF Receiver}

\vspace{0.2 cm}

For a ZF receiver, the detector matrix is defined as $\bm A = (\bm a_1, \bm a_2, \cdots, \bm a_K) = \widehat{\bm G} (\widehat{\bm G}^H \widehat{\bm G})^{-1}$. Clearly, for ZF receiver, $\bm A^H \widehat{\bm G} = \bm I_K$, i.e., $\bm a_k^H \widehat{\bm g}_q = \delta_{k,q} = 1$ if $k = q$ and $0$ if $k \neq q$, where $q = 1, 2, \ldots, K$ and $k = 1, 2, \ldots, K$. Substituting this result in \eqref{eq:modifyxhat}, we get $\text{ES}_k[t] = \sqrt{p_{\text{u}}} e^{-\sigma_{\omega_k}^2(t-(k-1))^2/2}\, x_k[t]$, where we have used the fact that $\E[e^{-j\Delta \omega_k (t - (k-1))}] = e^{-\sigma_{\omega_k}^2(t - (k-1))^2/2}$. Clearly, $\E[|\text{ES}_k[t]|^2] = p_{\text{u}} e^{-\sigma_{\omega_k}^2(t - (k-1))^2}$. Similarly $\E[|\text{SIF}_k[t]|^2]  =  p_{\text{u}} \left(1 - e^{-\sigma_{\omega_k}^2(t - (k-1))^2}\right)$, $\E[|\text{MUI}_k[t]|^2]  =  p_{\text{u}} \E \left[\{(\widehat{\bm G}^H\widehat{\bm G})^{-1}\}_{kk}\right] \sum\limits_{i=1}^{K}\frac{\beta_i \, \sigma^2}{K p_{\text{u}} \beta_i + \sigma^2}$, and $\E[|\text{EN}_k[t]|^2]  =  \sigma^2 \E \left[\{(\widehat{\bm G}^H\widehat{\bm G})^{-1}\}_{kk}\right]$.

\begin{lemma}
\label{mmseestresults}
\normalfont
With MMSE channel estimates, it can be shown that $\E \left[\{(\widehat{\bm G}^H\widehat{\bm G})^{-1}\}_{kk}\right] = \Big(\frac{1}{\beta_k} + \frac{\sigma^2}{K p_{\text{u}}\beta_k^2}\Big)/(M-K) $, where $\widehat{\bm G}$ is the MMSE estimate of effective channel gain matrix (see \eqref{eq:chanest}).
\end{lemma}


\begin{IEEEproof}
See Appendix A. 
\end{IEEEproof}

\vspace{-0.4 cm}

\begin{proposition}
\label{zfinfrate}
\normalfont
For the ZF receiver, the lower bound in \eqref{eq:infolb} is given by the achievable information rate $R_k^{\text{zf}}[t] = \log_2 (1 + \text{SINR}_k^{\text{zf}}[t])$, where

\vspace{-0.9 cm}

\small{\begin{IEEEeqnarray}{rCl}
\label{eq:zfsinrk}
\text{SINR}_k^{\text{zf}}[t] & = & \dfrac{e^{-\sigma_{\omega_k}^2(t - (k-1))^2}}{ \left[1 - e^{-\sigma_{\omega_k}^2(t - (k-1))^2}\right] + \dfrac{1}{M-K}\left( \dfrac{1}{\beta_k} + \dfrac{1}{K \beta_k^2 \gamma} \right) \left[\sum\limits_{i=1}^{K} \dfrac{\beta_i}{K \gamma \beta_i + 1} + \dfrac{1}{\gamma}\right] }.
\IEEEeqnarraynumspace
\end{IEEEeqnarray}}\normalsize

\vspace{-0.1 cm}

\noindent where $\gamma = \frac{p_{\text{u}}}{\sigma^2}$ and $t = K, K+1, \ldots, N_u -1$.
\end{proposition}

\vspace{-0.3 cm}

\begin{IEEEproof}
Using the expression for $\E \left[\{(\widehat{\bm G}^H\widehat{\bm G})^{-1}\}_{kk}\right]$ from Lemma~$1$, we get the expressions for $\E\Big[|\text{MUI}_k[t]|^2\Big]$ and $\E\Big[|\text{EN}_k[t]|^2\Big]$ (see paragraph before Lemma~$1$). Using expressions of $\E\Big[|\text{ES}_k[t]|^2\Big]$, $\E\Big[|\text{SIF}_k[t]|^2\Big]$, $\E\Big[|\text{MUI}_k[t]|^2\Big]$ and $\E\Big[|\text{EN}_k[t]|^2\Big]$ (see paragraph before Lemma~1) in the expression of $\text{SINR}_k[t]$ in \eqref{eq:infolb} we obtain \eqref{eq:zfsinrk}.
\end{IEEEproof}

\vspace{-0.7 cm}

\subsection{Mutual Information Analysis for the MRC Receiver}

For MRC receiver, $\bm A = \widehat{\bm G}$, or, $\bm a_k = \widehat{\bm g}_k,\, \forall k = 1, 2, \ldots, K$. Substituting this result in \eqref{eq:modifyxhat}, we get $S_k[t] = \sqrt{p_{\text{u}}} \, ||\widehat{\bm g}_k||^2 \, e^{-j \Delta \omega_k (t - (k-1))}$. The desired signal $\text{ES}_k[t]$ is therefore given by \small{$\text{ES}_k [t] = \E[S_k[t]]x_k[t] = \sqrt{p_{\text{u}}} \E[||\widehat{\bm g}_k ||^2]e^{-\sigma_{\omega_k}^2(t - (k-1))^2/2} x_k[t]$}\normalsize, or, \small{$\text{ES}_k [t] = \sqrt{p_{\text{u}}} \E[(\widehat{\bm G}^H\widehat{\bm G})_{kk}]e^{-\sigma_{\omega_k}^2(t - (k-1))^2/2}x_k[t]$}\normalsize. Therefore \small{$\E[|\text{ES}_k[t]|^2]  =  p_{\text{u}} \left(\E \Big[(\widehat{\bm G}^H\widehat{\bm G})_{kk}\Big]\right)^2 e^{-\sigma_{\omega_k}^2 (t - (k-1))^2}$}\normalsize. Similarly, \small{$\E[|\text{SIF}_k[t]|^2]  =  p_{\text{u}} \left[\E \Big[ \Big|(\widehat{\bm G}^H\widehat{\bm G})_{kk} \Big|^2\Big] - \left(\E \Big[(\widehat{\bm G}^H\widehat{\bm G})_{kk}\Big]\right)^2 e^{-\sigma_{\omega_k}^2 (t - (k-1))^2}\right]$}\normalsize, \small{$\E[|\text{MUI}_k[t]|^2]  =  p_{\text{u}} \left[ \sum\limits_{i=1, i\neq k}^{K} \E \Big[\Big|(\widehat{\bm G}^H\widehat{\bm G})_{ki}\Big|^2\Big]  + \sum\limits_{i=1}^{K} \frac{\beta_i \, \sigma^2}{K p_{\text{u}} \beta_i + \sigma^2} \E \Big[(\widehat{\bm G}^H\widehat{\bm G})_{kk}\Big] \right]$}\normalsize and \small{$\E[|\text{EN}_k[t]|^2]  =  \sigma^2 \E \Big[(\widehat{\bm G}^H\widehat{\bm G})_{kk}\Big]$}\normalsize.

\begin{lemma}
\label{mmseresults2}
\normalfont
With MMSE channel estimate $\widehat{\bm G}$ of the channel gain matrix $\bm G$ (see \eqref{eq:chanest}), it can be shown that $\E \left[(\widehat{\bm G}^H\widehat{\bm G})_{kk}\right] = M\frac{K p_{\text{u}}\beta_k^2}{K p_{\text{u}}\beta_k + \sigma^2}$, $\E \left[\Big|(\widehat{\bm G}^H\widehat{\bm G})_{kk} \Big|^2\right] = M(M+1)\left(\frac{K p_{\text{u}}\beta_k^2}{K p_{\text{u}}\beta_k + \sigma^2}\right)^2$ and, $\E \left[\Big|(\widehat{\bm G}^H\widehat{\bm G})_{ki} \Big|^2\right] = M\left(\frac{K p_{\text{u}}\beta_k^2}{K p_{\text{u}}\beta_k + \sigma^2}\right)\left(\frac{K p_{\text{u}}\beta_i^2}{K p_{\text{u}}\beta_i + \sigma^2}\right)$.
\end{lemma}

\begin{IEEEproof}
See Appendix B.
\end{IEEEproof}

\vspace{-0.5 cm}

\begin{proposition}
\label{mrcinfrate}
\normalfont
For the MRC receiver, the lower bound in \eqref{eq:infolb} is given by the achievable information rate $R_k^{\text{mrc}}[t] = \log_2 (1 + \text{SINR}_k^{\text{mrc}}[t])$, where $t = K, K+1, \ldots, N_u -1$, and

\vspace{-0.8 cm}

\small{\begin{IEEEeqnarray}{rCl}
\label{eq:mrcsinrk}
\text{SINR}_k^{\text{mrc}}[t] & = & \dfrac{e^{-\sigma_{\omega_k}^2(t - (k-1))^2}}{ \left[1 - e^{-\sigma_{\omega_k}^2(t - (k-1))^2}\right] + \dfrac{1}{M}\left( \dfrac{1}{\beta_k} + \dfrac{1}{K \beta_k^2 \gamma} \right) \left[\sum\limits_{i=1}^{K} \beta_i + \dfrac{1}{\gamma}\right] }.
\IEEEeqnarraynumspace
\end{IEEEeqnarray}}\normalsize

\end{proposition}

\vspace{-0.5 cm}

\begin{IEEEproof}
Firstly we substitute the expressions of \small{$\E \left[(\widehat{\bm G}^H\widehat{\bm G})_{kk}\right]$, $\E \left[\Big|(\widehat{\bm G}^H\widehat{\bm G})_{ki} \Big|^2\right]$ and $\E \left[\Big|(\widehat{\bm G}^H\widehat{\bm G})_{kk} \Big|^2\right]$}\normalsize from Lemma~\ref{mmseresults2} in the expressions for $\E\Big[|\text{ES}_k[t]|^2\Big]$, $\E\Big[|\text{SIF}_k[t]|^2\Big]$, $\E\Big[|\text{MUI}_k[t]|^2\Big]$ and $\E\Big[|\text{EN}_k[t]|^2\Big]$ (see paragraph before Lemma~$2$). Using these in \eqref{eq:infolb}, we get \eqref{eq:mrcsinrk}.
\end{IEEEproof}

\vspace{-0.2 cm}

\begin{theorem}
\label{arraygain}
\emph{(\textit{Achievable Array Gain})}
\normalfont
Consider $|\omega_k K| \ll \pi$, a fixed $K$, $N$ (length of pilot sequence) and a fixed desired information rate for the $t^{\text{th}}$ channel code of the $k^{\text{th}}$ user ($R_k^{\text{zf}}[t]$ and $R_k^{\text{mrc}}[t]$ defined in Proposition~\ref{zfinfrate} and Proposition~\ref{mrcinfrate} respectively). {For both the ZF and MRC receivers, as $M \to \infty$, the minimum required SNR $\gamma$ to achieve the fixed desired information rate decreases as $\frac{1}{\sqrt{M}}$. Alternatively, with $M \to \infty$ and $\gamma \propto 1/\sqrt{M}$, the achievable information rate for the $t^{\text{th}}$ channel code, i.e., $R_k^{\text{zf}}[t]$ or $R_k^{\text{mrc}}[t]$, approaches a constant value}.
\end{theorem}

\vspace{-0.3 cm}

\begin{IEEEproof}
We have observed from Remark $2$ that as $M \to \infty$ with $\gamma = \frac{c_0}{\sqrt{M}}$ (constant $c_0 > 0$), the MSE for CFO estimation converges to a constant limiting value, i.e., $\lim\limits_{M \to \infty, \gamma \, = \, \frac{c_0}{\sqrt{M}}} \sigma_{\omega_k}^2 = \zeta_0 > 0$ (constant). {Substituting this result in the expression for $\text{SINR}_k^{\text{zf}}[t]$ in \eqref{eq:zfsinrk} and also in the expression for $\text{SINR}_k^{\text{mrc}}[t]$ in \eqref{eq:mrcsinrk} with $\gamma = \frac{c_0}{\sqrt{M}}$ we have}

\vspace{-0.8 cm}

{\small{\begin{IEEEeqnarray}{rCl}
\label{eq:asymsinrkzf}
\lim\limits_{M \to \infty} \text{SINR}_k^{\text{mrc}}[t] \, = \, \lim\limits_{M \to \infty} \text{SINR}_k^{\text{zf}}[t] = \dfrac{e^{-\zeta_0 (t-(k-1))^2}}{1 - e^{-\zeta_0 (t - (k-1))^2} + \dfrac{1}{K \beta_k^2 c_0^2}} > 0 \,\, (\text{constant}).
\IEEEeqnarraynumspace
\end{IEEEeqnarray}}\normalsize}

\vspace{-0.6 cm}

\indent {From \eqref{eq:asymsinrkzf} it is clear that $R_k^{\text{zf}}[t] = \log_2 (1 + \text{SINR}_k^{\text{zf}}[t])$ and $R_k^{\text{mrc}}[t] = \log_2 (1 + \text{SINR}_k^{\text{mrc}}[t])$ would also approach constant limiting values as $M \to \infty$ with $\gamma \propto \frac{1}{\sqrt{M}}$}. 
\end{IEEEproof}

\begin{remark}
\normalfont
From Theorem $1$, it is clear that with every doubling in the number of BS antennas, the minimum required SNR to achieve a fixed per-user information rate decreases by approximately $1.5$ dB as long as the number of BS antennas $M$ is sufficiently large. {This shows that with the CFO estimation technique proposed in \cite{gcom2015}, the ZF receiver (also the MRC receiver) yields an $\mathcal{O}(\sqrt{M})$ array gain in the massive MIMO uplink. This is interesting since even for the ideal zero CFO scenario with ZF/ MRC receiver, the maximum achievable array gain is known to be only $\mathcal{O}(\sqrt{M})$ \cite{Ngo1}}. \hfill \qed
\end{remark}

\vspace{-0.4 cm}

\begin{remark}
\normalfont
From \eqref{eq:asymsinrkzf} we have \small{$\lim\limits_{M \to \infty} \text{SINR}_k^{\text{mrc}}[t] = \lim\limits_{M \to \infty} \text{SINR}_k^{\text{zf}}[t]$}\normalsize. Clearly, as $M \to \infty$ with $\gamma \propto 1/\sqrt{M}$, the achievable information rate for both the ZF and MRC receiver approach the same lower bound. This shows us the new result that even with CFO estimation/compensation, MRC and ZF receivers have the same performance when $M$ is sufficiently large. \hfill \qed
\end{remark}

\vspace{-0.8 cm}

\section{Numerical Results and Discussions}

\vspace{-0.2 cm}

We present a comparative discussion on the performance of the ZF and MRC receivers, with CFO estimation/compensation in frequency-flat massive MIMO uplink. For monte-carlo simulations, we assume an operating carrier frequency $f_c = 2$ GHz and a maximum CFO of $1$ PPM of $f_c$. The communication bandwidth is $B_c = 200$ KHz. The coherence interval and the maximum delay spread are $1$ ms and $5 \, \mu$s respectively. Thus $|\omega_k| \leq \frac{\pi}{50}$ and $N_c = 1 \text{ms}/ B_c = 200$ channel uses. The duration of uplink is $N_u = 100$ channel uses. The length of pilot sequence for CFO estimation is taken as $N = 100$ and the number of UTs is {$K = 10$}. At the start of each CFO estimation phase $\omega_k$ assumes a random value uniformly distributed in $[-\frac{\pi}{50}, \frac{\pi}{50}]$. Also for simplicity, we assume $\beta_k = 1$, $\forall k = 1, 2, \ldots, K$. The information rate for each user is also computed analytically using Propositions~$1$ and $2$ in \eqref{eq:sumrate} with $\sigma_{\omega_k}^2 = \E[(\widehat{\omega}_k - \omega_k)^2]$ replaced by its approximation in \eqref{eq:cfomse} with $G_k = 1$ (see Remark~$1$).

\begin{figure}[!t]
\vspace{-0.7 cm}
\centering
\includegraphics[width= 5 in, height= 2.43 in]{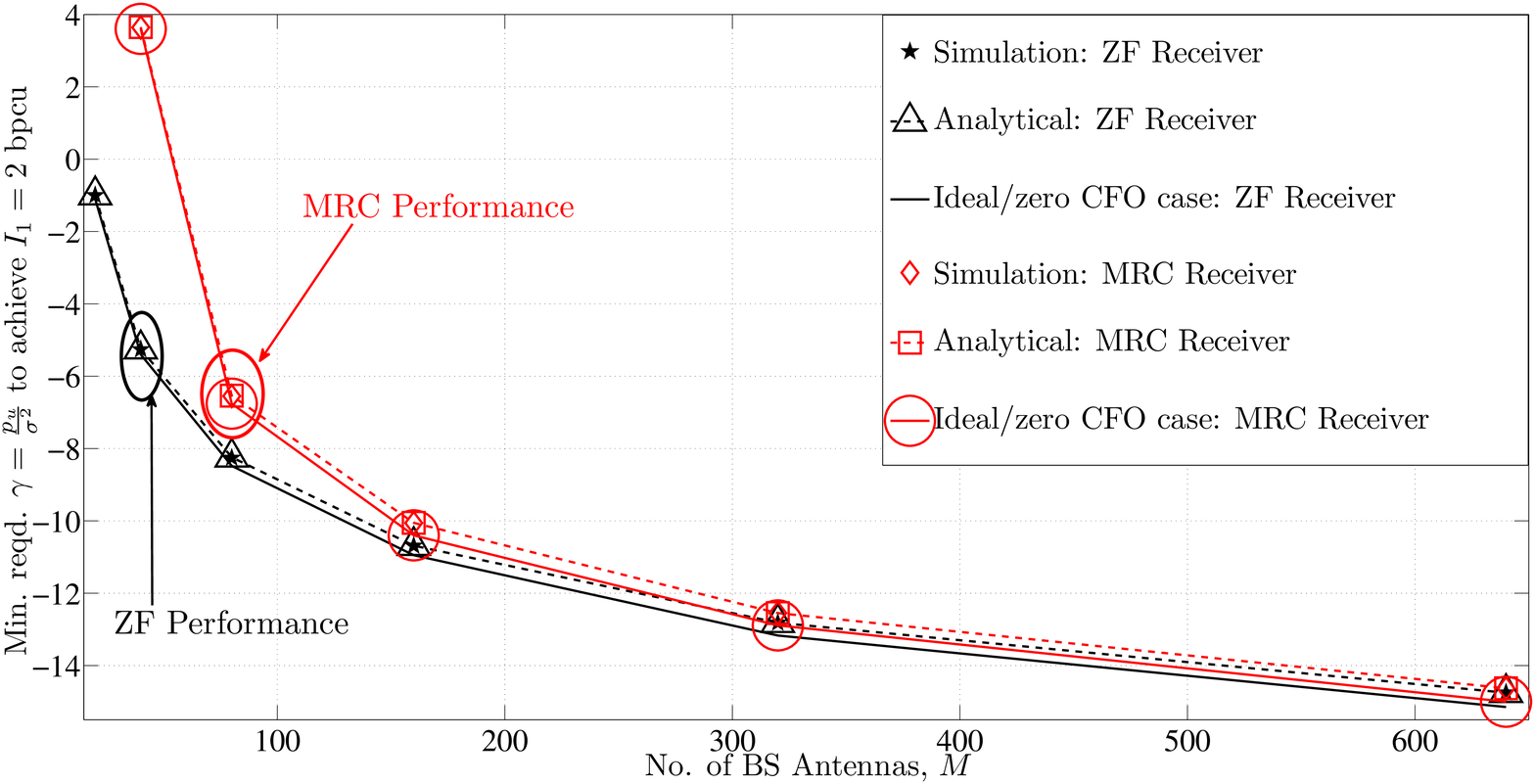}
\caption {{Plot of $\gamma = \dfrac{p_{\text{u}}}{\sigma^2}$ required to achieve $I_k = 2$ bpcu (for the first user ($k = 1$)) vs. $M$, fixed $K = 10$, $N = 100$.}}
\label{fig:arraygain}
\vspace{-0.7 cm}
\end{figure}

\begin{savenotes}
\begin{table}[b]
\caption[position=top]{{\textsc{SNR gap between ZF and MRC Receiver for fixed $M = 80$, $K = 10$.}}}
\label{table:snrgap}
\centering
\begin{tabular}{| c | c | c |}
\hline
Desired Per-User Information Rate & SNR gap for Ideal/zero CFO case & SNR gap with CFO compensation\\ 
\hline
\vspace{-0.6 cm} & \\
1 bpcu & {0.1} dB & {0.12} dB \\                 
\hline
\vspace{-0.6 cm} & \\
2 bpcu & {1.7} dB & {1.71} dB\\ 
\hline
\vspace{-0.6 cm} & \\
2.5 bpcu & {4.57} dB & {4.59} dB\\
\hline
\end{tabular}
\end{table}
\end{savenotes}

\par In Fig.~\ref{fig:arraygain} we plot the variation of the minimum required SNR $\gamma = p_{\text{u}}/\sigma^2$ (both analytical and simulated) to achieve a fixed information rate of $2$ bpcu (bits per channel use) for the 1\textsuperscript{st} user versus the number of BS antennas, $M$ (fixed {$K = 10$} and $N = 100$). Observe that the analytical approximation to the required $\gamma$ for both ZF and MRC is quite tight. Also for $M$ sufficiently large, with every doubling in $M$, the required $\gamma$ decreases roughly by $1.5$ dB (note the decrease in required SNR $\gamma$ from {$M = 320$ to $M = 640$}). This supports Theorem~$1$ and shows that with the discussed CFO estimation/compensation technique, an $\mathcal{O}(\sqrt{M})$ array gain is achievable. Also note that the required $\gamma$ for ZF and MRC is the same for sufficiently large $M >320$ (see Remark~$4$). However for finite $M$, ZF is more power efficient compared to MRC. {For example at $M = 80$, ZF requires approximately $1.7$ dB less power than MRC. Next we consider this extra SNR required by MRC when compared to ZF (denoted as SNR gap) for the same desired information rate $I_1 = 1, 2, 2.5$ bpcu (for the 1\textsuperscript{st} UT) for fixed $M = 80$ and $K = 10$ (see Table~\ref{table:snrgap})}. From Table~\ref{table:snrgap}, we make an interesting observation that the SNR gap between the ZF and MRC receivers is almost the same irrespective of whether we have the ideal zero CFO scenario or the residual CFO (after CFO compensation) scenario. Therefore the new result in this paper is that with CFO compensation, there is no significant degradation in the SNR gap when compared to the SNR gap in the ideal/zero CFO scenario.

\vspace{-0.5 cm}

\section{{Conclusion}}

{In this paper, we study the impact of low-complexity CFO estimation and compensation on the performance of ZF receiver in massive MIMO uplink in a flat fading environment and compare it to that of the MRC receiver. The tight closed-form analytical expressions for information rates of ZF and MRC reveal that an $\mathcal{O}(\sqrt{M})$ array gain is indeed achievable with CFO estimation. This is interesting since the best possible array gain for ideal zero CFO scenario is also known to be $\mathcal{O}(\sqrt{M})$. Finally the study of the SNR gap between ZF and MRC receivers for the same per-user information rate suggests that compared to the ideal zero CFO case, the performance degradation due to residual CFO is same for both the ZF and MRC receivers.}


\vspace{-0.5 cm}


\appendices
\vspace{-0.3 cm}

\section{Proof of Lemma 1}

\vspace{-0.3 cm}

From the relation $g_{mk} = h_{mk}\sqrt{\beta_k}$ we have $\bm G = \bm H \bm D^{1/2}$, where $\bm H \Define [h_{mk}]_{M \times K}$. Clearly from \eqref{eq:chanest} we have \small{$\widehat{\bm G}  =  (\sqrt{K p_{\text{u}}} \bm G \bm \Phi_0 + \bm N)\widetilde{\bm D} \mya  (\sqrt{K p_{\text{u}}} \bm H \underbrace{\bm D^{1/2} \bm \Phi_0}_{\Define \, \bm X} + \underbrace{\bm N \bm \Phi_0^H \bm D^{-1/2}}_{\Define \, \bm V} \bm D^{1/2} \bm \Phi_0)\widetilde{\bm D} = \Big(\underbrace{\sqrt{K p_{\text{u}}} \bm H + \bm V}_{\Define \, \bm Z} \Big)\bm X \widetilde{\bm D} = \bm Z \bm X \widetilde{\bm D}$}\normalsize, where $(a)$ follows from the fact that $\bm \Phi_0^H \bm \Phi_0 = \bm \Phi_0 \bm \Phi_0^H = \bm I_K$ and $\widetilde{\bm D} = (\sqrt{K p_{\text{u}}} \bm I_K + \frac{\sigma^2}{\sqrt{K p_{\text{u}}}}\bm D^{-1})^{-1}$. Let $\bm n_k$ and $\bm v_k$ be the $k^{\text{th}}$ columns of $\bm N$ and $\bm V$ respectively. Since $\bm n_k \sim \mathcal{C}\mathcal{N}(0, \sigma^2 \bm I_M), \,\, k = 1, 2, \cdots, K$ are all i.i.d. random vectors, $\bm v_k = (\bm \Phi_0^H \bm D^{-1/2})_{kk} \, \bm n_k,\,\, k = 1, 2, \cdots, K$, are also independently distributed as $\mathcal{C}\mathcal{N}(0, \frac{\sigma^2}{\beta_k}\bm I_M)$.

\par We also note that the columns of $\bm H$ and $\bm V$ are independent of each other. Clearly, the same is also true for the columns of $\bm Z$. Therefore we can write $\bm Z = \bm U \bm Q$, where $\bm Q \Define \left(K p_{\text{u}}\bm I_K + \sigma^2 \bm D^{-1}\right)^{1/2}$ and $\bm U \Define (\bm u_1, \bm u_2, \cdots, \bm u_K)$, where $\bm u_k \sim \mathcal{C}\mathcal{N}(0, \bm I_M) \,\, \forall k = 1, 2, \ldots, K$ are i.i.d. random vectors. Now, we have $\Big\{(\widehat{\bm G}^H\widehat{\bm G})^{-1}\Big\}_{kk}  = \{(\widetilde{\bm D}\bm \Phi_0^H \bm D^{1/2} \bm Q \bm U^H\bm U \bm Q \bm D^{1/2} \bm \Phi_0 \widetilde{\bm D})^{-1}\}_{kk}$, which follows from the fact that $\widehat{\bm G} = \bm Z \bm X \widetilde{\bm D}$, $\bm X = \bm D^{1/2} \bm \Phi_0$ and $\bm Z = \bm U \bm Q$. Since $\bm D$, $\bm \Phi_0$, $\bm Q$ and $\widetilde{\bm D}$ are all diagonal, $\Big\{(\widehat{\bm G}^H\widehat{\bm G})^{-1}\Big\}_{kk} =  \Big[\big|\bm T_{kk}\big|^2\Big]^{-1} (\bm W^{-1})_{kk}$, where $\bm T \Define \bm Q \bm D^{1/2} \bm \Phi_0 \widetilde{\bm D}$ and $\bm W \Define \bm U^H \bm U \sim \mathcal{W}_M(M, \bm I_M)$ is a $K \times K$ central Wishart matrix with $M$ degrees of freedom. 

\vspace{-0.9 cm}

\small{\begin{IEEEeqnarray}{rCl}
\label{eq:wishart1}
\nonumber \text{Clearly,} \, \, \E\Big[\Big\{(\widehat{\bm G}^H\widehat{\bm G})^{-1}\Big\}_{kk}\Big] & = & \Big[\big|\bm T_{kk}\big|^2\Big]^{-1} \E[(\bm W^{-1})_{kk}] = \left(\dfrac{K p_{\text{u}} \beta_k + \sigma^2}{K p_{\text{u}} \beta_k^2}\right)\frac{1}{K}\E \Big[tr(\bm W^{-1})\Big] \myb \dfrac{K p_{\text{u}} \beta_k + \sigma^2}{(M-K)K p_{\text{u}} \beta_k^2},
\IEEEeqnarraynumspace
\end{IEEEeqnarray}}\normalsize

\vspace{-0.2 cm}

\noindent where $(b)$ follows from $\E[tr(\bm W^{-1})] = \frac{K}{M-K}$ \cite{Tulino}.

\vspace{-0.5 cm}

\section{Proof of Lemma 2}

\vspace{-0.3 cm}

From Appendix~A we know that $\bm W = \bm U^H \bm U$ is a central Wishart matrix with $M$ degrees of freedom, i.e., $\bm W_{kk}$ is $\chi^2 (2M)$ (chi-squared) distributed. Therefore from definition of $\widehat{\bm G}$ we have

\vspace{-0.8 cm}

\small{\begin{IEEEeqnarray}{lCl}
\nonumber \E \Big[(\widehat{\bm G}^H\widehat{\bm G})_{kk}\Big]  \mya \big |\bm T_{kk}\big |^2 \E[\bm W_{kk}] = M \left(\dfrac{K p_{\text{u}} \beta_k^2}{K p_{\text{u}} \beta_k + \sigma^2}\right), \,\,\,  \E \left[\Big|(\widehat{\bm G}^H\widehat{\bm G})_{kk} \Big|^2\right]  =   M(M+1)\left(\frac{K p_{\text{u}}\beta_k^2}{K p_{\text{u}}\beta_k + \sigma^2}\right)^2, \,\,\, \text{and}\\
\label{eq:form2}
\nonumber \E \left[\Big |(\widehat{\bm G}^H \widehat{\bm G})_{ki} \Big |^2\right]  \myb  \E \left[ \Big | \bm T_{kk}^{\ast} \bm W_{ki} \bm T_{ii}\Big |^2\right] = \big |\bm T_{kk}\big |^2 \big |\bm T_{ii}\big |^2 \E \left[ \Big |\bm W_{ki}\Big|^2\right] \myc \left(\dfrac{K p_{\text{u}} \beta_k^2}{K p_{\text{u}} \beta_k + \sigma^2}\right)\left(\dfrac{K p_{\text{u}} \beta_i^2}{K p_{\text{u}} \beta_i + \sigma^2}\right) M,
\IEEEeqnarraynumspace
\end{IEEEeqnarray}}\normalsize

\vspace{-0.2 cm}

\noindent \noindent where $(a)$ and $(b)$ follow from the facts that $\widehat{\bm G} = \bm Z \bm X \widetilde{\bm D}$, $\bm X = \bm D^{1/2} \bm \Phi_0$, $\bm Z = \bm U \bm Q$ and $\bm T = \bm Q \bm D^{1/2} \bm \Phi_0 \widetilde{\bm D}$ (see Appendix A). Also $(c)$ follows from the fact that $\E[|\bm W_{ki}|^2] = \E[|\bm u_k^H \bm u_i|^2] = M$, since $\bm u_k, \, \forall \, k = 1, 2, \ldots, K$ are i.i.d. $\mathcal{C}\mathcal{N}(0, \bm I_M)$.

\vspace{-0.5 cm}


%

%


\ifCLASSOPTIONcaptionsoff
  \newpage
\fi



%


\bibliographystyle{IEEEtran}
\bibliography{IEEEabrvn,mybibn}


\end{document}